\documentstyle[12pt]{article}

\let\vev\VEV

\def\beq{\begin{equation}}
\def\eeq{\end{equation}}
\def\bea{\begin{eqnarray}}
\def\eea{\end{eqnarray}}

\relax
\begin{document}
\def\ra{\rightarrow}
\begin{titlepage}
\baselineskip 0.35cm
\begin{center}
\today
\hfill SNUTP 93-11\\
\null\hfill \\
\vskip 1.5cm
{\large \bf  Hadronic axion window and the big-bang
nucleosynthesis}
\vskip 1.7cm
{Sanghyeon Chang}
\vskip .05cm
{ Department of Physics and Center for Theoretical Physics}
\vskip .05cm
{Seoul National University, Seoul, 151-742 Korea}
\vskip 0.5cm
{Kiwoon Choi}
\vskip .05cm
{ Department of Physics, Chonbug National University}
\vskip 0.05cm
{Chonju, 560-756 Korea}

\vskip 1.5cm

{\bf Abstract}
\end{center}
\begin{quotation}
\baselineskip .73cm
Hadronic axions with the decay constant
 $f_a\simeq 10^{6}$ GeV
may fulfill all astrophysical and laboratory constraints discussed so far.
In this paper, we reexamine  the possibility of the hadronic axion
window while taking into account the uncertainties of
 some parameters describing low energy axion dynamics.
It is found that $f_a$ in the range from $3\times 10^5$ GeV
to $3\times 10^6$  can not be excluded by existing arguments.
We then examine the implication of this hadronic axion window for
the big-bang nucleosynthesis (NS) by evaluating
the energy density of thermal axions at the nucleosynthesis epoch.
Our analysis yields  $(\rho_a/\rho_{\nu})_{NS}=0.4\sim 0.5$ which
exceeds slightly the current best bound $(\rho_a/\rho_{\nu})_{NS}\leq 0.3$.

\end{quotation}
\baselineskip .75cm
\end{titlepage}
\setcounter{page}{2}

\noindent
{\bf 1. \, Introduction}

One of the most attractive  solutions to the strong CP problem
is to introduce a spontaneously
broken Peccei-Quinn  symmetry  which gives $\theta=0$
dynamically $\cite{strongcp}$.
This solution  predicts the existence of a pseudo-Goldstone
boson, the axion, whose decay constant $f_a$
is tightly constrained by
astrophysical and cosmological considerations $\cite{astro}$.
One frequently quoted window for
the axion decay constant satisfying  all phenomenological
constraints is  $f_a=10^{10}\sim 10^{12}$ GeV.
Besides this, for hadronic axions
which have  vanishing {\it tree level}   coupling to the electron,
there can be another  window
(the hadronic axion window) around $f_a\simeq 10^6$ GeV.
The existence of the hadronic axion window
relies upon (i)
the axion-electron coupling is radiatively induced and
thus  highly suppressed $\cite{kaplan}$,
(ii)  axions on the window  are  trapped inside the supernova core
 $\cite{turner}$,
and (iii) a significant
cancellation can occur
in the axion-photon coupling between the short distance
contribution (from the electromagnetic anomaly at $f_a$) and
the long distance contribution (from the axion-pion mixing at the QCD scale)
$\cite{kaplan}$.
Based upon these points, $f_a\simeq 10^6$ GeV can be consistent
with a variety of the strong  astrophysical constraints including
those from the supernova SN1987A $\cite{astro}$.

The  hadronic axion window can
be relevant for some models of dynamical symmetry
breaking. Then it would  become important
to examine the phenomenological viability of the window
more carefully.  To this end, recently the effect of
relic hadronic axions upon
the diffuse extragalactic background radiation
has been studied extensively $\cite{ressell}$.
By examining the photon flux from decaying axions
$\cite{photon}$,
it has been noted that
a severe cancellation should occur in axion-photon coupling
for $f_a\simeq 10^6$ GeV
to be consistent with the observed data $\cite{ressell}$.
In fact,  there is a possible  loophole for this conclusion.
Even when  $f_a$ and the axion-photon coupling are fixed,
there still exists a large uncertainty of the photon flux
associated with the uncertainty of
the light quark mass ratio $z=m_u/m_d$.
For a given value of $f_a$,
the axion mass is determined by  $z$ (see eq. (5)).
In ref. $\cite{ressell}$, $z=0.56$ was taken to arrive
at the above conclusion.
However it has been argued by several authors $\cite{qmass}$
that, due to instanton effects,  the true value of $z$ can  be
significantly smaller than the conventionally used
value 0.56. A smaller $z$  would give  a smaller axion
mass for which  both the position and the intensity
of the photon flux become lower.
Clearly then the required  cancellation
of the axion-photon coupling becomes  weaken.
At any rate, in generic hadronic axion models,
one can achieve the required cancellation by
appropriately arranging  the electromagnetic anomaly at
$f_a$.

In fact, the axion-photon coupling is the least known
parameter among those  describing
the low energy dynamics of hadronic axions.
It is thus desirable to consider an effect
which is independent of the  axion-photon coupling.
In this paper, we wish
to examine  one such effect,
that on the big-bang nucleosynthesis\footnote{
The constraints imposed by the nucleosynthesis
on the interactions of majorons have been
discussed by  Bertolini and Steigman
$\cite{bertolini}$}.
The organization of this paper is as follows.
In the next section, we  briefly  review low energy
axion couplings $\cite{kaplan,georgi}$ which will be used
in later discussion.
In sec 3  we collect arguments defining the hadronic axion
window. Taking into account the uncertainties of the
 involved  parameters, we argue that existing
arguments allow the window: $3\times10^5\leq f_a\leq 3\times 10^6$
(in GeV unit).
In sec 4 we  evaluate the rate of thermal axion production
after the quark-hadron phase transition which was roughly
estimated in ref. $\cite{ktbook}$.
This will determine the energy density
of thermal axions at the nucleosynthesis epoch.
For the window defined in sec 3,
we find   $(\rho_a/\rho_{\nu})_{NS}=0.4\sim 0.5$
which  exceeds slightly the current best bound
$(\rho_a/\rho_{\nu})\leq 0.3$ $\cite{olive}$.
Sec 5 is given for the conclusion.

\noindent
{\bf 2. \, Low energy couplings of hadronic axion}

In this section, we review the
low energy  couplings of hadronic axions
$\cite{kaplan,georgi}$.
We will mainly follow  ref. $\cite{georgi}$.
To be definite, let us consider  the
simplest hadronic axion model $\cite{kim}$
which was first considered by  Kim.
 The  model contains
an isosinglet Dirac quark $Q$ with the electromagnetic
charge  $Y_Q$
 and also a gauge singlet complex scalar field
$\sigma$.  These exotic fields
carry  nonzero PQ charges: 1, $-1$, and 2 for
$Q_L$, $Q_R$ and $\sigma$ respectively, while all other
fields have vanishing PQ charge.
The PQ symmetry is spontaneously broken
by $\vev{\sigma}=f_a/\sqrt{2}$ and then the axion $a$ can be defined by
\beq
\sigma=\frac{1}{\sqrt{2}}(f_a+\rho)\exp(i a/f_a).
\eeq
Integrating out heavy degrees of freedom,
viz $Q$ and $\rho$ whose masses
are of the order of $f_a$,  we obtain
the axion effective lagrangian  at the renormalization
point  $\mu=f_a$:
\beq
\frac{1}{2}(\partial_{\mu}a)^2+\frac{a}{f_a}(
\frac{g_c^2}{32\pi^2}G
\tilde{G}+\kappa\frac{g_1^2}{32\pi^2}Y\tilde{Y}).
\eeq
Here $G$ and $Y$ denote the field strengths
 of $SU(3)_c$ and $U(1)_Y$ respectively,
while $\tilde{G}$ and $\tilde{Y}$ are their duals, and
 $\kappa=6Y_Q^2$
denotes  the electromagnetic anomaly arising from
the triangle loop of  $Q$.
Note that there is no  tree level axion-electron
coupling,  which is the characteristic property of hadronic axions.

Given  the axion effective lagrangian at $\mu=f_a$,
we can derive  a variety of axion properties at low energies
by  scaling the effective lagrangian
 down to the QCD scale $\Lambda\simeq 1$ GeV $\cite{georgi}$.
Below $\Lambda$,  the QCD anomaly and instanton effects become strong
and also the chiral symmetry is spontaneously broken.
It is then convenient to eliminate the $aG\tilde{G}$ coupling
through an axion-dependent
chiral rotation of the light quark fields $q=(u, d, s)$:
\beq
q_L\ra \exp (iaQ_A/f_a) q_L, \quad q_R\ra \exp (-ia Q_A/f_a) q_R,
\eeq
where
\beq
Q_A=M^{-1}/2{\rm tr}(M^{-1}),
\eeq
and $M={\rm diag}(m_u,m_d,m_s)$ denotes the light quark mass
matrix.
With this chiral rotation, the axion field $a$ becomes
a physical mass eigenstate with
\beq
m_a=\frac{\sqrt{z}}{1+z}\frac{f_{\pi} m_{\pi}}{f_a},
\eeq
where  $f_{\pi}=93$ MeV denotes the pion decay constant and $z=m_u/m_d$.
Also the axion couplings
with the electron and the photon  renormalized at $\mu
<\Lambda$ are given by
\beq
C_{a\gamma}\frac{a}{f_a}\frac{e^2}{32\pi^2}F^{\mu\nu}\tilde{F}_{\mu\nu}
+C_{ae}\frac{\partial_{\mu} a}{f_a} \bar{e}\gamma^{\mu}\gamma_5 e,
\eeq
where\footnote{
Note that our result for $C_{ae}$ is smaller than the Srednicki's
in ref. $\cite{kaplan}$
 by the factor $1/2\pi$.}
\bea
&&C_{a\gamma}=\kappa-\frac{2(4+z)}{3(1+z)},
\nonumber \\
&&C_{ae}=\frac{3\alpha_{em}^2}{8\pi^2}[\, \kappa\ln(\frac{f_a}{\mu})-
\frac{2(4+z)}{3(1+z)}\ln(\frac{\Lambda}{\mu})].
\eea

There are two kind of axion couplings with hadrons; derivative and
nonderivative couplings $\cite{georgi}$.
In fact, axion phenomenology
associated with the hadronic axion window
is  governed mainly by derivative couplings.
At $\mu\simeq \Lambda$, after the chiral rotation (3),
axion  derivative couplings with the light quarks are given by
\beq
\frac{\partial_{\mu}a}{f_a}[ \,\frac{1}{6}
 (1+\gamma)\bar{q}\gamma^{\mu}\gamma_5 q
+\frac{1}{2}x_a\bar{q}\gamma^{\mu}\gamma_5\lambda^a q \,],
\eeq
where $x_a={\rm tr}(Q_A\lambda^a)$
and  $\gamma(\mu)$ is due to the renormalization of
the singlet  current  occurring between $f_a$ and $\mu$.
Here ${\rm tr}(\lambda^a\lambda^b)=2\delta_{ab}$.
This then gives
 \beq
 \frac{\partial_{\mu} a}{f_a}(\frac{1}{6}J^{\mu}_0+\frac{1}{2}
x_aJ^{\mu}_a),
\eeq
where $J^{\mu}_0$ and $J^{\mu}_a$ are the hadronic currents made of
the pseudoscalar meson octet
 $\Sigma=\exp (i\pi^a\lambda^a/f_{\pi})$ and the baryon octet
$B=B^a\lambda^a/\sqrt{2}$:
\bea
J^{\mu}_a&=&\frac{i}{2} f_{\pi}^2\, {\rm tr}(\lambda^a
(\Sigma D^{\mu}\Sigma^{\dagger}
  -\Sigma^{\dagger}D^{\mu}\Sigma))
  +{\rm tr}(\bar{B}\gamma^{\mu}[\lambda^a_P(\xi), B])
  \nonumber \\
 && +F\,{\rm tr}(\bar{B}\gamma^{\mu}\gamma_5[\lambda^a_S(\xi),
B])+D\,{\rm tr}(\bar{B}\gamma^{\mu}\gamma_5\{\lambda^a_S(\xi),B\}),
 \nonumber \\
J^{\mu}_0&=&S\, {\rm tr}(\bar{B}\gamma^{\mu}\gamma_5 B).
\eea
Here $\lambda^a_P(\xi)=
(\xi\lambda^a\xi^{\dagger}-\xi^{\dagger}\lambda^a\xi)/2$,
 $\lambda^a_S(\xi)=(\xi\lambda^a\xi^{\dagger}+
 \xi^{\dagger}\lambda^a\xi)/2$
 for $\xi^2=\Sigma$, and
 $F$ and $D$ denote the baryon matrix elements of
 the $SU(3)$ octet axial vector currents,  while $S$ is defined as
\beq
(1+\gamma)<B|\bar{q}\gamma^{\mu}\gamma_5 q|B>= S n^{\mu},
\eeq
where $n^{\mu}$ denotes the covariant spin vector
of the baryon state $B$.
Note that $S$ is independent of the QCD renormalization point
$\mu$.

Among axion couplings with hadrons, those of particular interest
for us are  the axion-nucleon couplings ${\cal L}_{aN}$
responsible for the  processes $N N\rightarrow NN a$ and $N\pi\rightarrow
N a$ and also the axion-pion couplings ${\cal L}_{a\pi}$
for $\pi\pi\rightarrow\pi a$.
These couplings can be read  from eqs. (9) and (10) giving
\bea
{\cal L}_{aN}&=&\frac{\partial_{\mu}a}{f_a}
[C_{an}\bar{n}\gamma^{\mu}\gamma_5 n+
C_{ap}\bar{p}\gamma^{\mu}\gamma_5p+
i\frac{C_{a\pi N}}{f_{\pi}}(\pi^+\bar{p}\gamma^{\mu}n-
\pi^{-}\bar{n}\gamma^{\mu} p)]
\nonumber \\
{\cal L}_{a\pi}&=& C_{a\pi}\frac{\partial_{\mu} a}{f_af_{\pi}}
(\pi^0\pi^+\partial_{\mu}\pi^-
+\pi^0\pi^-\partial_{\mu}\pi^+-2\pi^+\pi^-\partial_{\mu}\pi^0)
\eea
where
\bea
C_{an}&=&\frac{z}{2(1+z)}F+\frac{z-2}{6(1+z)}D+\frac{1}{6}S,
\nonumber \\
C_{ap}&=&\frac{1}{2(1+z)}F+\frac{1-2z}{6(1+z)}D+\frac{1}{6}S,
\nonumber \\
C_{a\pi N}&=&\frac{1-z}{2\sqrt{2}(1+z)}
\nonumber \\
C_{a\pi}&=&\frac{1-z}{3(1+z)}
\eea

Several parameters appear  in eqs. (7) and (13)
to determine
the low energy axion couplings for a given value of $f_a$.
Let us discuss the values of those parameters.
First of all, the electromagnetic anomaly
$\kappa$ is due to the physics at the axion scale
$f_a$ and thus is totally unknown.
In principle, one can arrange the particle content at $f_a$
to have  $\kappa$  taking an arbitrary value of order
unity. The  axial vector coupling constants  $F$ and  $D$ of baryons
can be determined by the nucleon and hyperon beta  decays
as $F= 0.47$ and $D=0.81$
$\cite{manohar}$.  Then recent
EMC measurement\footnote{
In fact, what the EMC measurement gives  is the value
of  $S/(1+\gamma)$  at $\mu\simeq 3$ GeV where
 $\gamma$ is negligibly small. Note that at leading order
 the evolution of $\gamma$ is given by
  $\mu d\gamma/d\mu=-3(\alpha_c/\pi)^2\gamma$
  $\cite{kodaira}$ with
$\gamma(\mu=f_a)=0$. It has been pointed out $\cite{kaplan1}$
that elastic neutral current experiment gives a similar
result $S\simeq 0.09\pm 0.24$.}
 of the polarized proton
structure function  gives
$ S\simeq 0.13\pm 0.2$ $\cite{manohar}$.
About $z$, one usually uses the result $z= 0.56$
from first order chiral
perturbation theory $\cite{weinberg}$.
In second order chiral perturbation theory,
$z$ receives a correction $\delta z=O(m_s/4\pi f_{\pi})$
which is essentially  due to  instanton effects $\cite{qmass}$.
A larger value of this instanton-induced correction
means a smaller $z$. It has been argued that $\delta z$
can be large enough to imply $z=0$ $\cite{qmass}$.
With this point taken into account, we will allow $z$ take
a value significantly smaller than 0.56.

\noindent
{\bf 3. \, The window}

In this section, we  will collect arguments  which define
the hadronic axion window.  Taking into
account the uncertainties of  $\kappa$ and $z$, we will
argue that existing arguments do {\it not}  exclude the window
\beq
3\times 10^5 \, {\rm GeV} \leq f_a \leq 3\times 10^6
\, {\rm GeV}.
\eeq

A key property of hadronic axion
is that its coupling to the electron
is radiatively induced and thus is highly suppressed.
As is well known, helium ignition of red giants
provides a strong constraint on the axion-electron
coupling $\cite{dearborn}$,
$C_{ae}m_e/f_a\leq  1.5\times 10^{-13}$
in our notation.
This gives\footnote{
Since our result for the radiatively induced
axion-electron coupling $C_{ae}$ is smaller than the Srednicki's,
the lower bound on $f_a$ is relaxed by the factor
$1/2\pi$. Note that the hadronic axion window becomes widen as a
result of this point.}
  $f_a\geq \kappa \times 10^5$ GeV
which is satisfied by the  window (14) for $\kappa$
of order unity.

The list of other arguments defining  the hadronic axion window
is as follows:
(A) the axion's effect on the neutrino burst from SN1987A
$\cite{turner,burrows}$,
(B) the  effect of axions emitted from SN1987A on the Kamiokande II
detector $\cite{engel}$,
(C) the effects of axion emission on the evolution
of helium burning low-mass stars $\cite{raffelt}$,
(D) the effect of decaying relic axions on the diffuse
extragalactic background radiation $\cite{ressell}$.
In regard to the argument (A),
axions on the window (14) would be trapped inside
the supernova core $\cite{turner}$. A detailed numerical analysis
for  this ``trapping regime"  has been performed
by Burrows et al. $\cite{burrows}$
who have concluded that if $f_a\leq 8C_{a N}\times 10^6$
GeV (here $C_{aN}\simeq (C_{an}^2+C_{ap}^2)^{1/2}/{\sqrt{2}}$)
axions would be so strongly trapped that
axion emission would {\it not} have a significant effect
on the neutrino burst\footnote{
Although less complete,
axion trapping was discussed also by Ishizuka and Yoshimura
$\cite{ishizuka}$ and Engel {\it et al.} $\cite{engel}$.}
Also by considering (B), it has been pointed out that
$f_a\leq 2C_{aN}\times 10^6$ GeV would have produced an unacceptably
large signal at the Kamiokande detector $\cite{engel}$.
These then give   the allowed window\footnote{Here
we ignore possible uncertainties of the astrophysical
arguments leading to this range of $f_a$.}
\beq
2  C_{aN}\times 10^6 \, {\rm GeV}\leq f_a \leq 8 C_{aN}\times
10^6 \, {\rm GeV}.
\eeq
Using $F=0.47$,  $D=0.81$, and $S=0.13\pm 0.2$, and allowing $z$
vary between zero and 0.56, we  find $C_{aN}=0.15\sim 0.4$
for which the above window goes into
that of eq. (14).

Can the window (14) be compatible  with the constraints from (C) and (D)?
The  arguments  (C) and (D)  provide limits
to the axion-photon coupling.
{}From (C), one has derived $\cite{raffelt}$
$C_{a\gamma}\leq (f_a/10^7 \, {\rm GeV})$.
It has been pointed out that,
for $z=m_u/m_d=0.56$ and  $f_a\simeq 10^6$ GeV, the argument (D)
can lead to a more stringent limit $\cite{ressell}$,
say $C_{a\gamma}\leq 0.02$.
However the validity of the latter limit  strongly
relies upon the used  value of $z$.
Since $m_a\propto \sqrt{z}$, if $|z|\ll 0.56$
which is an open possibility $\cite{qmass}$,
both the position and  the intensity of the photon-line
from decaying axions  would be significantly  lowered.
Clearly then the  limit from (D) becomes invalid or at least weaken.

For hadronic axions, $C_{a\gamma}$ can be divided into two
pieces, the short
distance contribution $\kappa$ denoting the electromagnetic
anomaly  at the axion scale $f_a$ and the long distance part
$-2(4+z)/3(1+z)$ arising from the axion-pion mixing.
Then an interesting point is there can be a severe
cancellation between the short and long distance contributions.
In order to see this,
let us consider two  examples:
(e.1) $z=0.56$, $Y_Q=4/7$; (e.2) $z=0.05$,
$Y_Q=2/3$.  Here $Y_Q$ denotes the electromagnetic
charge of the heavy Dirac quark $Q$ and thus $\kappa=6Y^2_Q$.
We then have (e.1) $C_{a\gamma}=O(10^{-2})$; (e.2)
$C_{a\gamma}\simeq 0.1$.
These examples simply show that a significant cancellation
can occur rather naturally, but
the degree of cancellation is highly model-dependent.
At any rate, it is  true that,
for $f_a$ on the window (14),
one can  always arrange the model  to have  $\kappa$
for which $C_{a\gamma}$ becomes  small enough
to satisfy the constraints from (C) and (D).
We thus conclude that, due to our ignorance
of  $\kappa$ ( and also of $z$),  the hadronic axion window (14)
is allowed also by (C) and (D).

\noindent
{\bf 4. Thermal hadronic axions and the nucleosynthesis}

In this section, we consider the implication of the hadronic axion
window (14) for the big-bang nucleosynthesis (NS).
As is well known, in the standard model of NS,
the energy density of exotic particles  at the NS epoch
is severely constrained.
The primordial yields of $^4$He is sensitive to
the universal expansion rate $H\propto \rho^{1/2}$
at $T\simeq 1$ MeV  and, thereby, leads to a
bound on any exotic contribution to the energy
density $\rho$.
The recent work of Walker et al. $\cite{olive}$
provides the most stringent
bound on the axion energy density normalized
to the energy density of the electron-neutrino:
\beq
\delta N_{\nu}\equiv \left(\frac{\rho_a}{\rho_{\nu}}\right)_{NS}\leq 0.3.
\eeq

Since both  axions and  neutrinos are relativistic at $T\simeq 1$ MeV,
one simply has
$\delta N_{\nu}=(4/7)(T_a/T_{\nu})^4_{NS}$.
Suppose axions were initially  in equilibrium with the thermal
bath of normal particles ($\gamma$, $e^{\pm}$, $\nu$, ...) at
high temperature, but later become decouple  at  temperature $T_D>1$
MeV.
Entropy conservation then leads to the relation
\beq
\delta N_{\nu} =\frac{4}{7}\left(\frac{43/4}{g_{\ast}(T_D)}\right)^{4/3},
\eeq
where $g_{\ast}$  counts  {\it interacting}
degrees of freedom  whose entropy density is given by
$s=2\pi^2g_{\ast} T_{\nu}^3/45$.
Here the axion decoupling temperature $T_D$ can be found
by comparing the expansion rate $H$ with the axion
production rate
\beq
\Gamma=n_a^{-1}\sum_{i,j} n_i n_j \vev{\sigma_{ij} v},
\eeq
where $n_a$ denotes the axion number density and
 the sum extends to all production
processes involving as initial states the particles $i$ and $j$
which are still in equilibrium at $T_D>1$ MeV.
As was  pointed out by Bertolini and Steigman $\cite{bertolini}$,
for the NS constraint (16) to be satisfied, axions (or
any light scalar with $m\ll 1$ MeV) must  decouple
before $T= 100$ MeV.

Clearly $\delta N_{\nu}\leq 4/7$.
If axions decouple before the quark-hadron phase
transition, we would have $\delta N_{\nu}\leq 0.06$
because of the relative ``heating" of neutrinos during
the quark-hadron phase transition. We are thus led
to consider the possibility for axions in thermal equilibrium after
the quark-hadron phase transition.
Then there are two types of processes dominantly
producing thermal axions: (a) $\pi^0\pi^{\pm}\ra a\,\pi^{\pm}$,
$\pi^+\pi^-\ra a\,\pi^0$, and
(b) $\pi^0 N\ra a N$ ($N=n$ or $p$), $\pi^+ n\ra a p$,
$\pi^- p\ra a n$ and their CP conjugates.
In fact, the rate of the process (b)
was roughly estimated in ref. $\cite{turner}$.
Here we will carefully evaluate the rates of both (a) and (b).

Let $\Gamma_{(a)}$ and $\Gamma_{(b)}$ denote the axion
production rate of (a) and (b) respectively.
Then a straightforward computation yields
\bea
\Gamma_{(a)}&=&\frac{3}{1024\pi^5}\frac{1}{f_a^2f_{\pi}^2}C_{a\pi}^2I_{(a)},
\nonumber \\
\Gamma_{(b)}&=&\frac{1}{6\pi^3}\frac{1}{f_a^2f_{\pi}^2}[g_A^2(5C_{an}^2+5C_{ap}^2+2
C_{an}C_{ap})+6C_{a\pi N}^2]I_{(b)},
\eea
where
\bea
I_{(a)}&=&n_a^{-1}T^8\int dx_1dx_2\frac{x_1^2x_2^2}{y_1y_2}
f(y_1)f(y_2) \int^{1}_{-1}
d\omega\frac{(s-m_{\pi}^2)^3(5s-2m_{\pi}^2)}{s^2T^4},
\nonumber \\
I_{(b)}&=&n_a^{-1}n_NT^5\int dx_1 x_1 y_1^3f(y_1).
\eea
Here $f(y)=1/(e^y-1)$ denotes the pion distribution function,
$n_a$ and $n_N$ are the axion and nucleon number densities
at equilibrium,
$x_i=|\vec{p}_i|/T$,  $y_i=E_i/T$ ($i=1,2$),
and $s=2(m_{\pi}^2+T^2(y_1y_2-x_1x_2\omega))$.
To derive $\Gamma_{(a)}$,
we have used the pion-nucleon coupling
$g_A\bar{N}\gamma^{\mu}\gamma_5\partial_{\mu}\tilde{\pi}N$
where $g_A=F+D$ and $\tilde{\pi}=\pi^i\tau^i/2$ ($\tau^i=$
the Pauli matrices).

By performing numerical analysis for $I_{(a)}$ and $I_{(b)}$,
one can see that $\Gamma_{(a)}$ dominates over
$\Gamma_{(b)}$ when $T\leq 150$ MeV.
Here to be definite, we have used $z=0.56$, $F=0.47$,
$D=0.81$, and $S=0.13$, but
the result is quite insensitive to the allowed
variations of these parameters.
 Using $\Gamma/ H\simeq 1$ at $T=T_D$, we find\footnote{
Using the rough
estimate of $\Gamma_{(b)}$ given in ref. $\cite{ktbook}$,
$T_D\simeq 50$ MeV was
found in ref. $\cite{ressell}$. Our careful
evaluation yields   $\Gamma_{(b)}$
which is significantly smaller than that of ref. $\cite{ktbook}$.
This is mainly due to the suppression caused by the
axion-nucleon coupling constants. As a result, if one uses our
$\Gamma_{(b)}$ while ignoring $\Gamma_{(a)}$, one would get
$T_D=70\sim 100$ MeV
for $f_a=3\times 10^5\sim 3\times 10^6$ GeV.
However as we have mentioned, around $T_D$,
the pion process (a) largely dominates
over the nucleon process (b), leading to a lower
$T_D$.}
$T_D=30\sim 50$ MeV  for the range of $f_a$
from $3\times 10^5$ GeV to $3\times 10^6$ GeV.
With the relation (17), this gives
\beq
\delta N_{\nu}=0.4\sim 0.5,
\eeq
which is in slight contradiction to the NS bound (16).

\noindent
{\bf 5. Conclusion}

Hadronic axions with the decay constant around $10^6$ GeV
can escape from all astrophysical and laboratory
constraints discussed so far.  In this paper,
we have collected  arguments which define the hadronic axion
window and found that existing arguments allow
the window $3\times 10^5\leq f_a\leq 3\times 10^6$
(in GeV unit). The phenomenological viability
of this window relies strongly upon the model-dependent
cancellation that occurs in the axion-photon coupling.
As we have noted, in generic hadronic axion models,
it is not so unnatural to achieve the required cancellation.
Since the axion-photon coupling is quite model-dependent
and is totally unknown, it is desirable to consider an effect
which is independent of the axion-photon coupling.
We thus have considered the effect on the nucleosynthesis
which is determined by the axion-hadron couplings.
By evaluating the axion production rate after the quark-hadron
phase transition,
we could determine
the energy density of thermal axions at the nucleosynthesis
epoch. It is quite certain that $T_D< 100$ MeV and thus
the resulting  $\delta N_{\nu}$
exceeds  the lower bound $0.3$ $\cite{olive}$.
However since the deviation is not so significant, $\delta N_{\nu}=0.4\sim
0.5$,
it would still be difficult  to conclude that the hadronic axion window
is excluded by the primordial abundance of $^4$He.

\vskip 1.5cm
\noindent{\bf Acknowledgment}

This work is supported in part by  KOSEF through
CTP at Seoul National University and also by
the Ministry of Education.

\vfill\eject


\begin{thebibliography}{99}


\bibitem{strongcp}
J. E. Kim,
Phys. Rep.  150 (1987) 1; H. Y. Cheng,
Phys. Rep.  158 (1988) 1; R. D. Peccei, in: CP Violation,
ed. C. Jarlskog (World Scientific, Singapore, 1989).

\bibitem{astro}
M. S. Turner, Phys. Rep.  197 (1990) 67;
G. G. Raffelt, Phys. Rep.  198 (1990) 1.



\bibitem{kaplan}
D. B.  Kaplan, Nucl. Phys.  B260 (1985) 215; M. Srednicki,
{\it ibid}.  B260 (1985) 689.

\bibitem{turner}
M. S. Turner, Phys. Rev. Lett.  60 (1988) 1797.

\bibitem{ressell}
M. T. Ressell, Phys. Rev.  D44 (1991) 3001.

\bibitem{photon}
T. W. Kephart and T. J. Weiler, Phys. Rev. Lett.
 58 (1987) 171; M. S. Turner, Phys. Rev. Lett.
 59 (1987) 2489.

\bibitem{qmass}
H. Georgi and I. N. McArthur, Report No. HUTP-81/A011, 1981
(unpublished);
D. B. Kaplan and A. V. Manohar, Phys. Rev. Lett.  56
(1986) 2004; K. Choi, C. W. Kim and W. K. Sze, Phys. Rev. Lett.  61
(1988) 794; K. Choi, Nucl. Phys.   B383 (1992) 58.


\bibitem{bertolini}
S. Bertolini and G. Steigman, Nucl. Phys.  B387
(1992) 193.


\bibitem{georgi}
H. Georgi, D. B. Kaplan  and L. Randall, Phys. Lett.  B169
(1986) 73.

\bibitem{ktbook}
E. W. Kolb and M. S. Turner, {\it The Early Universe}
(Addison-Wesley, 1990), pp. 422-36.



\bibitem{olive}
K. A. Olive {\it et al}., Phys. Lett.  B236 (1990) 454;
T. P. Walker {\it et al}., Ap. J.  376 (1991) 51.

\bibitem{kim}
J. E. Kim, Phys. Rev. Lett.  43 (1979) 103;
M. A. Shifman, A. I. Vainshtein and V. I. Zakharov,
Nucl. Phys.  B166 (1980) 493.



\bibitem{manohar}
R. L. Jaffe and A. V. Manohar, Nucl. Phys.  B337 (1990) 509

\bibitem{kaplan1}
D. Kaplan  and A. Manohar, Nucl. Phys.  B310
(1988) 527.


\bibitem{kodaira}
J. Kodaira, Nucl. Phys.  B165 (1979) 129.

\bibitem{weinberg}
S. Weinberg, Trans. N. Y. Acad. Sci.  38 (1977) 185;
J. Gasser and H. Leutwyler, Phys. Rep.  87 (1982) 77.

\bibitem{dearborn}
D. Dearborn, D. Schramm and G. Steigman, Phys. Rev. Lett.
 56 (1986) 26.



\bibitem{burrows}
A. Burrows, M. T. Ressell and M. S. Turner, Phys. Rev. D42
(1990) 3297.



\bibitem{engel}
J. Engel, D. Seckel and A. C. Hayes, Phys. Rev. Lett.
 65 (1990) 960.


\bibitem{raffelt}
G. Raffelt and D. Dearborn, Phys. Rev.  D36 (1987) 2211.

\bibitem{ishizuka}
N. Ishizuka and M. Yoshimura,
Prog. Theor. Phys.  84 (1990) 233


\end{thebibliography}
\end{document}